\documentclass[twocolumn,showpacs,prt,aps]{revtex4-1}
\usepackage{dcolumn}
\usepackage{amssymb}

\draft

\newcommand{\eo}{e^{L_0t}\,}

\newcommand{\ej}{e^{L_jt}\,}

\newcommand{\eol}{e^{(L_0+L_1)\,t}\,}

\newcommand{\eot}{e^{(L_0+L_2)\,t}\,}

\newcommand{\eolt}{e^{(L_0+L_1+L_2)\,t}\,}

\newcommand{\eos}{e^{(L_0+L_1+\dots +L_s)\,t}\,}

\begin{document}

\title{Brownian particle in ideal gas:
explicit density expansions, \\ %
conditional probabilities, 
and 
amusing  properties of molecular chaos}

\author{Yu. E. Kuzovlev}

\affiliation{Donetsk
Physics and Technology Institute of NASU, %
ul. R.Luxemburg 72, 83114 Donetsk, Ukraine, %
\,e-mail:\, kuzovlev@fti.dn.ua}


\begin{abstract}
Explicit density expansions of non-equilibrium %
probability distribution functions for %
molecular Brownian particle in ideal gas are obtained %
in original form what visually implies (is %
exact solution to) the previously established dynamical virial %
relations. Role of these relations in unbiased analysis of %
molecular chaos properties in many-particle statistical %
mechanics, including the mobility %
1/f noise, is newly investigated %
in clear terms of conditional probabilities and averages.
\end{abstract}


\maketitle


\section{Introduction}

Here, we continue consideration %
of special but principally important problem of %
non-equilibrium statistical mechanics:\, factual statistics of %
random walk of molecular-size Brownian particle (BP) %
interacting with atoms of thermodynamically equilibrium %
ideal gas \cite{p0806,tmf,ig,p1105}. %
Namely, using a formal trick like %
suggested in our recent work \cite{p1203}, %
we shall derive explicit series expansions of the BP's %
and many-particle probability distribution functions %
over mean density of the gas. %
We shall see that these expansions, being represented %
in terms of the irreducible correlation %
(cumulant) functions, give unique formal solution %
to the exact ``virial relations''  %
\cite{p0806,tmf,ig,p0802,p0803} %
(if considered as equations for the correlation functions). %
Then we shall discuss some important consequences of these results, %
first of all, concerning existence of 1/f\,-type %
BP's mobility fluctuations and essentially non-Gaussian %
long-range statistics of BP's path (dramatic failure of %
the ``law of large numbers'' in true many-particle %
statistical mechanics).

\section{Density expansion of probability
distribution functions}

Let $\,N\gg 1\,$ ideal gas atoms are contained, along with %
the BP, in a box with volume\, %
$\,\Omega = N/n \gtrsim a^3N\,$\, %
where $\,n\,$ and  $\,a\,$\, %
denote mean density of the gas (concentration of atoms) %
and characteristic radius of BP-atom interaction, respectively. %
The interaction potential, $\,\Phi(\rho)\,$, %
is assumed repulsive and short-range. %
Let at initial time moment,  $\,t=0\,$, %
full normalized $\,(N+1)$-particle distribution function (DF) %
of this system is
\begin{eqnarray}
D_N(t=0)\,=\, F_0^{in}(R,V) \prod_{j=1}^N \frac %
{E(r_j-R)\,G_m(v_j)}{\int_{\Omega} E(r-R)\, d^3r} %
\,\,\, \,\, \label{iic}
\end{eqnarray}
Here\, $\,R\,,\,V\,$ and $\,r_j\,,\,v_j\,$ are coordinates and %
velocities of BP and atoms,\, %
$\,F_0^{in}=F_0(t=0)\,$ is BP's initial probability  %
distribution (normalized to unit),\, %

$\,G_m(v)=(2\pi T/m)^{-\,3/2}\exp{(- mv^2/2T)}\,$\, %
is Maxwell velocity distribution for particle with mass\, $\,m\,$,\, %
and  $\,E(\rho)=\exp{[-\Phi(\rho)/T]}\,$. %

Clearly, if $\,F_0^{in}(R,V)= W^{in}(R)\,G_M(V)\,$ %
(with $\,M\,$ being BP's mass) %
then the distribution (\ref{iic}), %
$\,D_N(t=0)\,\propto\, W^{in}(R)\,\exp{(-H^{(N)}/T)}\,$  %
(with $\,H^{(N)}\,$ being full system's Hamiltonian), represents %
thermodynamically equilibrium system,  %
but it is not statistically equilibrium (stationary) %
because of its spatial inhomogeneity in respect to %
BP's position if described by more or less localized %
initial distribution $\,W^{in}(R)\,$,\, %
e.g.\, $\,W^{in}(R)=\delta(R)\,$\, (which will be %
taken in mind below).

At $\,t>0\,$,\, %
$\,D_N(t)=\,\exp{(L^{(N)}t)}\,D_N(t=0)\,$, %
where $\,L^{(N)}\,$ denotes full system's Liouville operator %
(corresponding to $\,H^{(N)}\,$ and %
taking into account interactions with the box walls). %
For our system we can write
\begin{eqnarray}
L^{(N)}\,=\, L_0\,+\sum_{j=1}^N\,L_j\,\,, \label{il}
\end{eqnarray}
where operator\, $\,L_0\,$ represents BP itself, %
while operators\, $\,L_j\,$ with $\,j>0\,$ %
represents $\,j\,$-th atom itself  %
plus its mutual interaction with BP. %

Further, introduce sequence of $\,(s+1)$-particle %
($\,s=0,1,2\,\dots\,N\,$) marginal probability distribution %
functions (DF) for BP and $\,s\,$ atoms:
\begin{eqnarray}
F_s(t)\,=\,\Omega^s \int_{s+1} \dots \int_N %
D_N(t)\,=\, \label{ie}\\
=\, \Omega^s \int_{s+1} \dots \int_N  %
e^{\,L^{(N)}t}\,D_N(t=0)\,\,,  \nonumber
\end{eqnarray}
where\, $\,\int_j\,$\, symbolizes integration over %
variables of $\,j\,$-th atom. %
Thus, in respect to atoms' coordinates, %
these DF are ``normalized to volume'', %
i.e. defined standardly, like in the mentioned %
references and originally by Bogolyubov in \cite{bog}. %
Consider them under the thermodynamic limit, - when %
$\,N\rightarrow\infty\,$ and $\,\Omega\rightarrow\infty\,$ %
(boundaries of the container disappear at infinity) under %
fixed $\,N/\Omega = n=\,$const\,, - as functions of the %
mean gas density $\,n\,$. %

With this purpose, it is convenient, firstly, to introduce, %
like in \cite{p1203}, ``coherent product'' \,$\,\circ\,$\, %
of operator exponentials by definition as follows: %
\begin{eqnarray}
e^{A_1}\circ\,\dots\,\circ e^{A_s}\,\equiv\, %
e^{A_1+\dots + A_s}\, \,\nonumber
\end{eqnarray}

\begin{widetext}

Then, secondly, to use identities
\begin{eqnarray}
e^{\,L^{(N)}t}\, \,=\, \eo \prod_{j=1}^N %
\circ [1+(\ej -1)]\, 
\,=\,\eo +\,\eo \sum_{k=1}^N \, %
\sum_{\,1\leq j_1<\dots < j_k\leq N} \, %
\circ [e^{L_{j_1}t}-1\,] \dots \circ %
[e^{L_{j_k}t}-1\,]\,\, %
\label{iid}
\end{eqnarray}
and %
\begin{eqnarray}
e^{\,L^{(N)}t}= \eos \prod_{j=s+1}^N %
\circ [1+(\ej -1)] =\,\, \,\,\,\,\label{iids}\\ %
= \,\eos \{\,1+ \sum_{k=1}^{N-s}  %
\sum_{\,s+1\leq j_1<\dots < j_k\leq N\,} \, %
\circ [e^{L_{j_1}t}-1\,] %
\dots \circ [e^{L_{j_k}t}-1\,]\,\}\,\, %
\nonumber
\end{eqnarray}

Substituting this expansions to (\ref{ie}) %
we see that any finite-number term of corresponding series expansions %
of $\,F_s(t)\,$ is well defined in the thermodynamic limit. %
Combining Eqs.\ref{iic}, \ref{ie} and \ref{iids}, %
after quite obvious algebra, one obtains
\begin{eqnarray}
F_0(t,R,V)\,=\, {\bf S}(t)\, F_0^{in}(R,V)\, \,, \label{is}
\end{eqnarray}
where the BP's ``propagator'' $\,{\bf S}(t)\,$ is presented by series %

\begin{eqnarray}
{\bf S}(t)\,=\, \eo\,+ \eo \! \sum_{k=1}^\infty %
\frac {n^k}{k!} \prod_{j=1}^k \int_j %
\circ [\,\ej -1\,]\,g(x_j)\,= \, \label{ip}\\
=\,\eo +\,n\int_1 [\,\eol -\eo]\, g(x_1)\, %
+ \, \nonumber\\
\,+\, \frac {n^2}2 \int_1\int_2 %
[\,\eolt-\eol - \, \eot +\eo\,]\, %
g(x_1)\,g(x_2) \,+\, \dots \,\,, \nonumber
\end{eqnarray}
with\,\, $\,x_j\equiv \{\rho_j,v_j\}\,$,\, %
$\,\rho_j\equiv r_j-R\,$,\, %
$\,g(x)\,\equiv\, \exp{[-\Phi(\rho)/T]} \,G_m(v)\,%
=\, E(\rho)\,G_m(v)\,$,\, and\, %
$\,\int_j\,\dots =\int d^3\rho_j \int d^3v_j\,\dots\,$. %

At that, after change of variables from $\,r_j\,$ to %
$\,\rho_j =r_j-R\,$ and transition to the thermodynamic limit, %
the components of full Liouville operator (\ref{il}) can be chosen as %
\begin{eqnarray}
L_0\,=\,-V\nabla\,-\, f\,\nabla_P\,\,,
\label{il0}\\
L_j\,=\,(V-v_j)\,\nabla_j\,+\, %
\Phi^{\,\prime} (\rho_j)\,\left(\nabla_{p_j}\, %
-\, \nabla_P\right)\,\,, \label{ilj}
\end{eqnarray}
where\,  $\,\Phi^{\,\prime} (\rho)= %
\partial \Phi(\rho)/\partial \rho\,$,\, %
$\,\nabla =\partial/\partial R\,$,\, %
$\,\nabla_j =\partial/\partial \rho_j\,$,\, %
$\,\nabla_P =\partial/\partial P\,$,\, %
$\,\nabla_{p_j} =\partial/\partial p_j\,$,\, %
$\,P=MV\,$ and $\,p_j=mv_j\,$ are momenta of BP and atoms,\, %
and we introduced external force\, $\,f\,$\, acting onto BP. %

Notice that $\,k$-th term of (\ref{ip}) %
represents all such BP's trajectories what include %
$\,\geq k\,$ collisions with $\,k\,$ atoms, %
at least one collision with any of them. %
Indeed, if BP was not interacting with $\,j\,$-th of them,  %
i.e. $\,L_j\,$ was replaced by $\,(V-v_j)\nabla_j\,$, %
then integration $\,\int_j\,$ would turn into zero any %
whole $\,k\,$-th term of (\ref{ip}) with $\,k\geq j\,$. %

Similarly, from Eqs.\ref{iic}, \ref{ie} and %
\ref{iids} it follows that %
\begin{eqnarray}
F_s(t)\,=\,\eos\, \left\{\, %
1\,+ \sum_{k=1}^\infty %
\frac {n^k}{k!} \prod_{j=s+1}^{s+k} \int_j %
\circ [\,\ej -1\,]\,g(x_j)\,\right\}\, \,%
F_0^{in}\prod_{j=1}^s\, g(x_j) \,=\, %
\, \nonumber\\
=\,\eos\,F_s^{in}\,+ \eos \sum_{k=1}^\infty %
\frac {n^k}{k!} \,\left\{\, \prod_{j=s+1}^{s+k} \int_j %
\circ [\,\ej -1\,]\, \right\}\,\,F_{s+k}^{in} \,\,, %
\,\, \label{iss}\\
F_s^{in}\, \equiv \,F_0^{in} %
\prod_{j=1}^s\, g(x_j)\,\,\nonumber
\end{eqnarray}

Formulae (\ref{is})-(\ref{ip}) and (\ref{iss}) %
give explicit formal series expansions of non-equilibrium %
(time-dependent) DF in respect to mean gas density.

\section{Cumulant distribution functions and dynamical virial relations}

In addition to the DF $\,F_s(t)=F_s(t,R,V,x_1 \dots x_s)\,$, %
let us consider functions %
$\,C_s(t)=C_s(t,R,V,x_1 \dots x_s)\,$ %
defined by $\,C_0\equiv F_0\,$ and, %
at $\,s>0\,$, the series as  follow:

\begin{eqnarray}
C_s(t)\,=\,\eo \prod_{j=1}^{s} %
\circ [\,\ej -1\,]\,g(x_j)\, \left\{\, %
1\,+ \sum_{k=1}^\infty %
\frac {n^k}{k!} \prod_{j=s+1}^{s+k} \int_j %
\circ [\,\ej -1\,]\,g(x_j)\, \right\}\,F_0^{in} \,=\, %
\nonumber\\
\,=\, \sum_{k=0}^\infty \frac {n^k}{k!} %
\int_1 \dots\int_k \eo \prod_{j=1}^{s+k} %
\circ [\,\ej -1\,]\,g(x_j)\,\,F_0^{in} \,\,\, \label{cf}
\end{eqnarray}
(at $\,k=0\,$, of course, there is no integration).

From this definition it is clear that, first, %
$\,C_s(t=0)=0\,$ at $\,s>0\,$. %

Second,
\begin{eqnarray}
F_1(x_1)\,=\,F_0\,g(x_1) + C_1(x_1)\,\,, \, \label{fc}\\
F_2(x_1,x_2)\,=\,F_0\,g(x_1)\,g(x_1) + %
C_1(x_1)\,g(x_2) + g(x_1)\,C_1(x_2) + %
C_2(x_1,x_2)\,\,, \nonumber
\end{eqnarray}
and so on, where for simplicity %
only atoms' arguments are exposed. %
Hence, $\,C_s\,$ represent irreducible $\,(s+1)\,$-order %
correlations between BP and $\,s\,$ atoms and can be %
named ``cumulant functions'' (CF). %

Third, considering them as functions of the mean %
gas density too,\, $\,C_s=C_s(t,R,V,x_1 \dots x_s;\,n)\,$ %
(and correspondingly $\,F_s=F_s(t,R,V,x_1 \dots x_s;\,n)\,$), %
we can write
\begin{eqnarray}
\frac {\partial C_s}{\partial n}\,=\, %
\int_{s+1} C_{s+1} \,\, \label{dvr}
\end{eqnarray}
That are the ``virial relations'' which for the first %
time were found in an original way %
in \cite{p0802,p0803}  for  BP in (generally non-ideal and dense) %
fluids and then, in \cite{p0804} and \cite{p0806,tmf,ig}, %
derived directly from the BBGKY equations %
(in some of these references the CFs were designated as $\,V_s\,$). %
In essence, more reasonable name for such kind %
of results may be ``dynamical virial relations'' (DVR) \cite{p1203}. %

\,\,\,

\,\,\,

\end{widetext}
\section{Conditional probabilities and averages,
and DVR in their terms}

The DVR establish important connections %
between density dependence %
(and hence time and space dependence) of %
the BP's probability distribution, $\,F_0\,$, %
and degree and character of BP-gas statistical correlations. %
The first of relations (\ref{dvr}) already was carefully %
discussed in \cite{p0802,p0803}, \cite{p0804}, \cite{p0806,tmf,ig,p1105}, %
but here we shall refresh it and make more transparent %
by exploiting the useful concepts of conditional %
probabilities and conditional averages. %

Namely, let us introduce conditional two-particle DF: %
\begin{eqnarray}
\frac {F_1 (t,R,V,x;n)} %
{F_0(t,R,V;n)}\,=\, \frac %
{\overline{n}(t,x\,|R,V;n)}n\,\,,\,\,
\label{cdf1}
\end{eqnarray}
where\, $\,\overline{n}(t,x\,|R,V;n)= %
\overline{n}(t,\rho,v\,|R,V;n)\,$ %
is conditional mean value of  %
instant local density (concentration) %
of atoms in the $\,\mu\,$-space %
under given position and velocity of BP. %
That is
\begin{eqnarray}
\overline{n}(t,x\,|R,V;n)\,=\, %
\langle\, \widetilde{n}(t,x)\, %
\rangle_{R,V,n}\,\, \label{mr}
\end{eqnarray}
with %
\begin{eqnarray}
\widetilde{n}(t,x)\,=\sum_j\, %
\delta(x-x_j(t))\,= \, \, \label{iden}\\
=\sum_j\, \delta(\rho -(r_j(t)-R(t)))\, %
\delta(v-v_j(t))\,\, \nonumber
\end{eqnarray}
and $\,r_j(t),\, v_j(t)\,$, $\,R(t),\,V(t)\,$ being %
instant values of the system's variables. %
The condition is reflected by the subscript of %
angle brackets which in turn symbolize, %
as usually,  averaging over given %
statistical ensemble (determined by $\,F^{in}_s\,$). %

In terms of conditional statistical %
characteristics the first of DVR (\ref{dvr}), %
after its multiplication by $\,n\,$, %
clearly takes form
\begin{eqnarray}
\frac {\partial \ln\, F_0(t,R,V;n)} %
{\partial \ln\,n}\,=\, \nonumber\\ %
=\int [\, \overline{n}(t,x\,|R,V;n) - %
n\,g(x)]\, dx\,=\,\, \label{cdvr1}\\ %
= \int [\, \overline{n}(t,\rho\,|R,V;n) - %
n\,E(\rho)]\, d^3\rho\,\,, \nonumber
\end{eqnarray}
where we introduced also instant local gas density %
in configurational space,
\begin{eqnarray}
\widetilde{n}(t,\rho)\,=\, %
\int \widetilde{n}(t,x)\,d^3v\,=\sum_j\, %
\delta(\rho-\rho_j(t))\, 
\,, \, \, \label{in}
\end{eqnarray}
and\,
$\,\overline{n}(t,\rho|R,V,n) =$ $\langle\, \widetilde{n}(t,\rho)\, %
\rangle_{R,V,n}\,$\, as its conditional ensemble average value. %

If we are interested mainly in spatial %
dependence of $\,F_0=C_0\,$, i.e. in the %
BP's ``diffusion law''
\[
W(t,R;n)\,=\, \int C_0(t,R,V;n)\, d^3V\,\,,\,\,
\]
then, evidently, instead of (\ref{cdvr1}) %
we can write \cite{bkn}
\begin{eqnarray}
\frac {\partial \ln\, W(t,R;n)} %
{\partial \ln\,n}\,=\,\,\nonumber\\ %
=\int [\, \overline{n}(t,\rho\,|R;n)\, -\, %
n E(\rho)\,]\, d^3\rho\,= \,\nonumber\\ %
\,=\, \overline{\Delta N}(t\,|R,n)\,\,, %
\, \label{wdvr1} %
\end{eqnarray}
with $\,\overline{n}(t,\rho\,|R;n) = %
\langle\, \widetilde{n}(t,\rho)\,\rangle_{R,n}\,$ %
being conditional mean value of instant gas density at %
distance $\,\rho\,$ from BP under given BP's position, %
and $\,\overline{\Delta N}(t|R,n)\,$ %
corresponding conditional mean change of total number
of atoms in BP's surroundings.

Notice that the factor  $\,n\,E(\rho)\,$ is nothing but %
value of the  $\,\overline{n}(t,\rho\,|R;n)  \,$ %
under complete statistical equilibrium, when there are no %
statistical correlations between atoms and BP except those %
coming from their direct dynamic interaction. %
Therefore, integral in (\ref{wdvr1}),
$\,\overline{\Delta N}(t|R,n)\,$, %
characterizes specifically non-equilibrium ``excess'' %
(``historical'' \cite{tmf}) BP-gas correlations. %

\section{Diffusion law and fight of intuitions}

Let us consider the BP's diffusion law,  %
$\,W(t,R;n)\,$, assuming that \,(i) the BP-atom interaction %
potential is repulsive and short-range, %
\,(ii) the external force is absent, %
 $\,f=0\,$, i.e. our system all time stays in equilibrium, %
and \,(iii) $\,t\gg \tau\,$,\, where %
\,$\,\tau\,$\, is BP's mean free-flight time %
(or velocity relaxation time), %
\,$\,\tau\sim (\lambda/V_0) \sqrt{1+M/m}\,$\,, with %
$\,\lambda\,$\, being BP's mean free path and %
and \,$\,V_0=\sqrt{T/M}\,$\, characteristic %
thermal velocity. %
According to standard reasonings of gas kinetics %
and probability theory, firstly,  %
$\,\lambda=(\pi a^2n)^{-1}\,$ with $\,a\,$ denoting %
characteristic radius of BP-atom interaction. %
Secondly, $\,W(t,R;n)\,$ has Gaussian %
long-time asymptotic,
\begin{eqnarray}
W(t,R;n)\,\rightarrow \,
W^{(G)}_D(t,R) =\frac {\exp{(-R^2/4Dt)}} %
{(4\pi Dt)^{3/2}}\,\,, \,\, \,\, \label{g0}
\end{eqnarray}
where $\,D=D(n)\,$ is BP's diffusivity,\, %
$\,D(n)\sim V_0^2\tau \sim %
V_0\lambda \sqrt{1+M/m} \propto 1/n\,$\, for %
not too dense gas (at $\,4\pi a^3 n/3 \lesssim 1\,$). %

This asymptotic expresses the celebrated %
Bernoulli's ``law of large numbers'' \cite{jb} %
and is an integral part if common intuition %
cultivated by probability-theoretical way of thinking in %
statistical mechanics. %
However, it never was deduced from rigorous %
statistical mechanics in itself, without %
``art of conjectures'' (see %
comments and references in \cite{tmf,ig,p0802,p1203,i1,i2,p1}). %
Now, let us combine it with rigorous relation (\ref{wdvr1}) %
and with the inverse proportionality %
$\,D(n) \propto 1/n\,$. %
This yields

\begin{eqnarray}
\frac {\partial \ln\, W^{(G)}_{D(n)}} %
{\partial \ln\,n}\,= %
\frac 32 - \frac {R^2}{4Dt} =\,
\overline{\Delta N}(t|R,n)\,\,, %
\, \label{gdvr1} %
\end{eqnarray}

\ which, in turn, implies that
$\,\overline{\Delta N}(t|R,n)\,$ %
can take arbitrary large negative values. %
In other words, even in spite of equilibrium nature %
of the BP's diffusion, statistical correlations between gas %
and BP can involve arbitrary many atoms. %

This conclusion, however, is in contradiction to %
another common intuitive notion. Namely, that of %
Boltzmannian %
``molecular chaos'' which usually serves %
as logical reason of the ``law of large numbers''. %
Thus, the DVR visually reveal falsity %
of both these notions.

To avoid the contradiction, we should refuse hypothesis %
that rigorous statistical mechanics always %
allows realization of  ``molecular chaos'' and %
the ``law of large numbers'' %
in transport phenomena. %
And, hence, we should accept profound conjecture that BP-gas correlations %
have substantial but bounded value. That is %
the quantity $\,\overline{\Delta N}(t|R,n)\,$, - %
which can be named e.g. ``correlation number'', - %
satisfies
\begin{eqnarray}
-\,\Delta N_-\, \leq\, %
\overline{\Delta N}(t|R,n)\,\leq\,  %
\Delta N_+\,\,, \, \, \label{neq} %
\end{eqnarray}
where $\,\Delta N_-\,$ and $\,\Delta N_+\,$ %
are positive quantities independent %
(under long-range asymptotic) on $\,R\,$ %
and $\,t\,$ and, expectedly (at $\,a^3n\ll 1\,$), also on $\,n\,$. %

This means \cite{tmf,p0802,p0803} that instead of (\ref{g0}) %
we have to write, asymptotically,
\begin{eqnarray}
W(t,R;n)\,\rightarrow \,
\frac {\Psi(R^2/4Dt)} %
{(4\pi Dt)^{3/2}}\, %
\Xi(R^2/V_0^2t^2)\, \,\, \label{ng0}
\end{eqnarray}
with $\,D=D(n)\propto 1/n\,$, where, %
because of (\ref{wdvr1}) and (\ref{neq}), %
function $\,\Psi(z)\,$ possesses power-law tail: %
\begin{eqnarray}
\Psi(z)\,\propto \,\frac 1{z^\alpha}\,\,\, %
(z\gg 1)\,\,, \,\,\,\, \alpha = \frac 32 + %
\Delta N_-\,\,, \,\, \label{a}
\end{eqnarray}
and\, $\,\Xi(0)=1\,$. %
The cut-off function $\,\Xi(\cdot)\,$ %
(fast decreasing at infinity) can be omitted except %
when calculating higher-order statistical moments of $\,R\,$. %
Since at that, evidently,  %
$\, \Delta N_+ =\max_z\, \partial \ln\,W/ %
\partial \ln\,n\,=3/2\,$, the exponent $\,\alpha\,$ %
can be expressed as %
$\,\alpha =\Delta N_+ + \Delta N_- \,$. %
Explicit example of such (realistic) ``diffusion law'', - %
for probe atom of slightly non-ideal gas in the role of BP, - %
with the characteristic ``correlation number'' %
$\,\Delta N_- =2\,$,\, is presented in \cite{p1}. %
Thus, probabilities of large deviations of BP's path %
from its typical values ($\,|R|\sim \sqrt{6Dt}\,$) %
are giant as compared  with predictions of  %
the (idealistic) Gaussian law.

In  \cite{tmf,p0802,p0803}, these consequences %
of the first of DVR (\ref{dvr}) were considered in %
different fashion, in terms of characteristic %
volume, $\,\Omega_c\,$, occupied by BP-gas correlations %
in the $\,\rho\,$-space. %
It appears in the above formulae if we notice \cite{bkn} that, %
due to non-negativity of $\,\overline{n}(t,\rho\,|R,n)\,$, equality
(\ref{wdvr1}) implies inequality
\begin{eqnarray}
\frac {\partial \ln\, W(t,R;n)} %
{\partial \ln\,n}\,\gtrsim \, %
-\,n\,\Omega_c \,\, \label{om} %
\end{eqnarray}
(we neglect contributions $\,\sim a^3\,$ %
in comparison with $\,\Omega_c\,$). %
This coincides with (\ref{neq}) at %
$\,\Omega_c = \Delta N_-/n\,$. %
An unambiguous formal definition of $\,\Omega_c\,$ %
was done in \cite{tmf,p0802,p0803}.

The asymptotic (\ref{ng0})-(\ref{a}) means that BP has no %
\,{\it a priori\,} certain diffusivity. %
Instead, a variety of possible BP's trajectories is  %
characterized by some effective distribution, %
$\,U(D\,;t,n)\,$, of its diffusivity, so that
\begin{eqnarray}
W(t,R;n)\,\rightarrow
\int_0^\infty  W^{(G)}_{D^\prime}(t,R)\,\, %
U(D^\prime\,;t,n)\, dD^\prime\,\,\, \,\label{ng}
\end{eqnarray}
According to (\ref{ng0}) and (\ref{a}), the %
distribution must have definite power-law %
tail and thus look e.g. like
\begin{eqnarray}
U(D\,;t,n) \propto %
\frac {D_0^{\Delta N_-}} %
{D^{\Delta N_- +1}}\, %
\exp{\left(-\frac {D_0}D\right)}\, %
\Xi\left(\frac D{V_0^2 t}\right)\, %
\,\, \label{a1}
\end{eqnarray}
with $\,D_0=D_0(n)\propto 1/n\,$ %
being characteristic diffusivity scale %
and $\,\Xi(\cdot )\,$ nearly the same cut-off function as %
in (\ref{ng0}). Currespondingly, in (\ref{ng0})
\begin{eqnarray}
\Psi(z) \,=\, \frac {\Gamma(\alpha)} %
{\Gamma(\alpha -3/2)} \cdot \frac 1{(1+z)^{\,\alpha}}\,\, \, \label{a2}
\end{eqnarray}
with\, $\,\alpha = \Delta N_- +3'2\,$. %
Coefficient $\Gamma(\alpha)/ %
\Gamma(\alpha -3/2)$ is normalizing factor %
ensuring that (asymptotically) %
$\,\int W(t,R;n)\, d^3R\,=1\,$. %
This is direct analogue of %
distributions found in \cite{p1} (for probe atom of weakly non-ideal %
gas) and in \cite{p2} (for BP different from non-ideal gas atoms \cite{fn}) %
in the ``collisional approximation'' of BBGKY equations %
\cite{i1,i2,p1,p2}. %

According to the universal ``generalized fluctuation-dissipation %
relations'' (see e.g. \cite{bkn} and \cite{i1,i2,bk3} %
and references therein), the representation (\ref{ng})-(\ref{a1}), %
with %
\begin{eqnarray}
W^{(G)}_D(t,R) =\frac %
{\exp{[-(R-Df/T)^2/4Dt\,]}} %
{(4\pi Dt)^{3/2}}\, \,\, \label{gf}
\end{eqnarray}
in place of (\ref{g0}), is valid also for non-equilibrium %
Brownian motion under non-zero external force, $\,f\neq 0\,$, %
at least for sufficiently small one, $\,f\lambda\ll T\,$. %

The resulting distribution predicts and describes 1/f-type %
fluctuations of BP's mobility $\,\mu=D/T\,$, as well as %
at $\,f=0\,$ similar 1/f-type fluctuations of BP's diffusivity %
\cite{tmf,p1105,p0802,p0803,bkn,i1,i2,p1,bk3,yuk, %
ffnp,pr157,bk12,p1008,eiphg,p1207}, %
thus confirming logics of the Krylov's ``art of conjectures'' %
\cite{kr}.

At $\,D_0(f/T)^2t\gg 1\,$ (when drift contribution to BP's path %
$\,R(t)\,$ exceeds diffusive contribution) this mobility/diffusivity %
1/f\,-noise visually manifests itself in probability distribution %
of  $\,R(t)\,$'s projection onto $\,f\,$'s direction %
\cite{p0802}: according to (\ref{ng}) and (\ref{gf}), %
its shape nearly reproduces that of $\,U(D\,;t,n)\,$ %
(analogous distributions of charge carriers' mobilities %
in electronics were detected by time-of-flight measurements  \cite{p1008}).

\section{Higher-order conditional %
probabilities and virial relations}

It is not hard to see that expansions (\ref{iss}) and (\ref{cf}) or,  %
equivalently, formulae (\ref{fc}) together with the DVR (\ref{dvr}) %
imply relations
\begin{eqnarray}
\frac {\partial F_s }{\partial n}\,= %
\int_{s+1} [\,F_{s+1} -g(x_{s+1})\,F_s\,]\, %
\,\, \label{fdvr} %
\end{eqnarray}
After dividing by $\,F_s\,$ and multiplying by $\,n\,$ %
this yields many-particle analogue of (\ref{cdvr1}), %
\begin{eqnarray}
\frac {\partial \ln\, F_s} {\partial \ln\,n}\,=\,%
\overline{\Delta N}_s(t|R,V,x^{(s)},n)\,=\, \label{cdvr}\\ %
\,=\,\int [\,\overline{n}(t,x\,|R,V,x^{(s)}) - %
g(x)\,n\,]\, dx \,\, \nonumber %
\end{eqnarray}
with $\,x^{(s)}=\{x_1\dots x_s\}\,$ and %
$\,\overline{n}(t,x\,|R,V,x^{(s)},n)\,$ being conditional %
average of $\,\widetilde{n}(t,x)\,$ under condition that %
BP is in given state and, besides, some $\,s\,$ atoms %
already occupy $\,s\,$ exactly known states. %

Then, introducing  many-particle DF in configurational space, %
\[
W_s(t,R,\rho^{(s)};n)= %
\int F_s\, d^3V\,d^3v_1\dots d^3v_s\,\,,
\]
where $\,\rho^{(s)}=\{\rho_1\dots \rho_s \}\,$, %
one can transform (\ref{fdvr})-(\ref{cdvr}) to
\begin{eqnarray}
\frac {\partial \ln\, W_s} {\partial \ln\,n}\,=\, %
\overline{\Delta N}_s(t|R,\rho^{(s)},n)\,=\,\, \label{wdvr}\\ %
=\,\int [\,\overline{n}(t,\rho\,|R,\rho^{(s)},n) - %
E(\rho)\,n\,]\,\, d^3\rho \,\, \nonumber %
\end{eqnarray}
with $\,\overline{n}(t,\rho\,|R,\rho^{(s)},n) = %
\langle \,\widetilde{n}(t,\rho)\,\rangle %
_{R,\rho^{(s)},n}\,$ %
representing average under condition that  $\,s\,$
atoms are known already to be at given distances from BP. %

To interpret these formulae correctly, recall that, - by  %
statistical definition of DFs $\,F_s\,$ (see also \cite{bog} and below), - %
different arguments $\,x_j\,$ of $\,F_s\,$, relate to different atoms. %
Therefore,  $\,s\,$ atoms entering %
into the conditions in fact do not enter into %
$ \,\widetilde{n}(t,x)\,$ and $\,\widetilde{n}(t,\rho)\,$. %
As the consequence, a contribution they give to %
$\,\overline{\Delta N}(t|R,n)\,$ in (\ref{wdvr1}) is absent %
in $\,\overline{\Delta N}_s(t|R,\rho^{(s)},n)\,$ in (\ref{wdvr}), %
so that we can write
\begin{eqnarray}
\overline{\Delta N}_s(t|R,\rho^{(s)},n)\,\geq \, %
\overline{\Delta N}(t|R,n)\,-\,s\,\, \label{dn}
\end{eqnarray}
At that, of course, %
$\,\overline{\Delta N}_s(t|R,\rho^{(s)},n)\,\rightarrow $ %
$\overline{\Delta N}(t|R,n)\,$ when points %
$\,\rho^{(s)}\,$ are far from the ``correlation volume'' $\,\Omega_c\,$, %
i.e. $\,\rho^{(s)}\rightarrow\infty\,$, and equality %
in (\ref{dn}) can be achieved only at small emough  %
$\,\rho^{(s)}\,$ belonging to $\,\Omega_c\,$. %

Together with these reasonings, formulae %
(\ref{wdvr}), (\ref{dn}) and (\ref{neq}) naturally do prompt %
that for $\,s>0\,$, %
- in addition to (\ref{neq}), (\ref{ng0}) and (\ref{a2}), - %
the following asymptotical estimates take place (at $\,f=0\,$): %
\begin{eqnarray}
\min\, %
\overline{\Delta N}_s(t|R,\rho^{(s)},n)\, %
=\, -\Delta N_- \,-\,s\,\,, \,\,\label{neqs}\\
%
\overline{W}_s(t,R;n)\,\rightarrow \,
\frac {\Psi_s(R^2/4Dt)} {(4\pi Dt)^{3/2}}\, %
\,\Xi(R^2/V_0^2t^2) \ \,,\, \label{ngs}\\
\Psi_s(z)\,=\, \frac {\Gamma(\alpha_s)} %
{\Gamma(\alpha_s-3/2)\, (1+z)^{\alpha_s}} \,\propto \, %
\frac {\Psi(z)}{(1+z)^s}\,\,,\,\, \label{as}\\
\alpha_s\, =\, \Delta N_- + s  +3/2\,\,,\, \label{ass}
\end{eqnarray}
where the minimum is taken over $\,R\,$ and $\,\rho^{(s)}\,$, %
and $\,\overline{W}_s(t,R;n)\,$ means result of suitable smoothing %
of $\,W_s(t,R,\rho^{(s)}\,;n)\,$ over small enough %
$\,\rho^{(s)}\,$ from the ``correlation volume''. %
At that, $\,\int \overline{W}_s(t,R;n)\, %
d^3R\,=1\,$, as the above basic definition of $\, W_s\,$ does require.  %

Thus, the sequence of functions $\,\overline{W}_s(t,R;n)\,$ %
qualitatively coincides with sequence %
$\,\overline{W}_s(t,R)/(1+sm/M)\,$ from \cite{p2} (see also \cite{p0802}), %
and coincidence would be also quantitative %
if the ``correlation number'' %
$\,\Delta N_-\,$ was equal to $\,1+M/m\,$ (see below). %
This observation confirms that exact DVR (\ref{fdvr})-(\ref{wdvr}) %
and the ``collisional approximation'' \cite{i1,i2,p1,p2} %
are rather close approaches. %

One can notice that the sequence 
$\,\overline{W}_s =\overline{W}_s(t,R;n)\,$ %
satisfies (at $\,D\propto 1/n\,$) simple recurrent relations %
\begin{eqnarray}
\overline{W}_{s+1}= %
\left[\,1+\,\frac n{\Delta N_- +s}\, %
\frac {\partial}{\partial n}\,\right]\, %
\overline{W}_s \,\,\, \label{rr}
\end{eqnarray}
(more generally, $\,n\,\partial/\partial n\,$ may be reolaced %
by $\,-D\,\partial/\partial D\,$). %
Iterating them, after some combinatoric algebra one can obtain
\begin{eqnarray}
\overline{W}_s =\prod_{k=0}^{s-1} %
\frac {\Delta N_- +s+\,n \,\partial/\partial n} %
{\Delta N_- +s}\,\,W_0\,= \nonumber\\ = %
\sum_{k=0}^s \left( \begin{array}{c} %
s \\ k \end{array} \right)\, %
\frac{\Gamma(\Delta N_-)} %
{\Gamma(\Delta N_- +k)}\, %
\,n^k\,\frac {\partial^k}{\partial n^k}\,\, W_0\,\,  \label{ir}
\end{eqnarray}
Right-hand side here is mere result of %
``normal ordering'' of multiplication and differentiation %
operators $\,n\,$ and $\,\partial/\partial n\,$ from left side. %

\,\,\,

At\, $\,f\neq 0\,$ (in the linear drift response regime), %
correspondingly, %

\begin{widetext}

\begin{eqnarray}
\overline{W}_s\, %
\rightarrow\, \label{ngsf}
\int_0^\infty  W^{(G)}_{D^\prime} %
(t,R-D^\prime ft/T)\,\, %
U_s(D^\prime;t,n)\, dD^\prime\,\,,\, \\
U_s(D;t,n)\,=\, %
\frac {D_0^{\Delta N_- +s}} %
{\Gamma(\Delta N_- +s)\, D^{\Delta N_- +1+s}}\, %
\exp{\left(-\frac {D_0}D\right)}\, %
\Xi\left(\frac D{V_0^2 t}\right)\, %
\,,\, \label{asf}
\end{eqnarray}
where effective diffusivity distributions\, %
$\,U_s\,$ obey recurrent relations
\begin{eqnarray}
U_{s+1}(D;t,n)\,=\, %
\left[\,1+\,\frac n{\Delta N_- +s}\, %
\frac {\partial}{\partial n}\,\right]\, %
U_s(D;t,n)\, \, \label{rrf}\\
\end{eqnarray}
at\, $\,D_0\propto 1/n\,$.

\section{Generating functionals and virial relations}

For many purposes it is convenient to accumulate all DFs %
in a single generating functional (GF),
\begin{eqnarray}
\mathcal{F}\{t,R,V,\psi\,;\,n\,\}\,=\, %
F_0\,+  \label{gf} 
\sum_{s=1}^\infty\, \frac {n^s}{s!} %
\int_1 \! \dots \! \int_s F_s\,\, %
\psi(x_1) \dots \psi(x_s)\, \, 
\end{eqnarray}
with $\,\psi(x)\,$ being formally arbitrary probe function. %
Following the original general Bogolyubov's construction of such %
functionals in \cite{bog} (where probe function %
$\,u(x)=n\psi(x)\,$ was used instead of $\,\psi(x)\,$), %
we can introduce our one as thermodynamic limit of  %
\begin{eqnarray}
\mathcal{F}_N\{t,R,V,\psi;n\}\,=\, %
F_0\,+ \sum_{s=1}^N\, %
\frac {N!}{s!(N-s)!\Omega^s} %
\int_1\dots\int_s F_s\,\, %
\psi(x_1) \dots \psi(x_s)\, \,\equiv\, \, \nonumber\\ \,\equiv
\int_1\! \dots\! \int_N D_N(t) \prod_{j=1}^N\, %
[\,1\,+\,\psi(x_j)\,]\,\,=\, \label{gfn} 
\int_1\dots\int_N \exp{\left\{ %
\sum_{j=1}^N\, %
\ln\,[1+\psi(x_j)]\,\right\}}\, %
D_N(t)\, 
\,=\, \nonumber\\ \,=\, %
\int_1\dots\int_N \exp{\left\{ %
\int \widetilde{n}(t,x)\, \ln\,[1+\psi(x)]\, %
dx\, \right\}}\, %
D_N(t)\,\,\,,   \nonumber
\end{eqnarray}
where $\,\Omega =N/n\,$,\, and we involved the ``microscopic gas density''\, %
$\, \widetilde{n}(t,x)=\sum_{j=1}^N \delta(x-x_j(t))\,$\, whose %
statistics is determined by the total DF %
$\,D_N(t)\,$ in turn  determined by (\ref{iic})  and (\ref{ie}). %
The latter expressions highlight statistical meaning of %
of the functional (\ref{gf}), %
$\,\mathcal{F}=\lim\,\mathcal{F}_N\,$, %
allowing us to write \cite{p1105}\,:
\begin{eqnarray}
\mathcal{F}\{t,R,V,\psi;n\}\,=\, %
\left\langle\,\delta(R-R(t))\, \delta(V-V(t))\,\, %
\exp{\left\{ %
\int \widetilde{n}(t,x)\, \ln\,[1+\psi(x)]\, %
dx\, \right\}}\, \right\rangle_n \,= \, %
\,\nonumber\\ \,=\, %
F_0(t,R,V;n)\, \left\langle\, %
\exp{\left\{ %
\int \widetilde{n}(t,x)\, \ln\,[1+\psi(x)]\, %
dx\, \right\}}\, \right %
\rangle_{R,V,n} \,\, \,\, \label{gfc}
\end{eqnarray}
Here, as above,\, $\,\langle \dots \rangle_{R,V,n}\,$\, %
is symbol of conditional averaging under given %
BP's variables, so that the latter angle brackets %
represent conditional characteristic functional %
of random field $\,\widetilde{n}(t,x)\,$, with %
$\,\xi(x) =\ln\,[1+\psi(x)]\,$ in the role of probe function. %

Next, introduce, in full analogy with (\ref{gf}), %
GF of the CFs (\ref{cf}): %
\begin{eqnarray}
\mathcal{C}\{t,R,V,\psi\,;\,n\,\}\,=\, %
C_0\,+  \label{gc} %
\sum_{s=1}^\infty\, \frac {n^s}{s!} %
\int_1 \! \dots \! \int_s C_s\,\, %
\psi(x_1) \dots \psi(x_s)\, \,=\, %
\mathcal{P}\{t,R,V,\,(1+\psi)\,n\,\}\,\,, \, \\
\mathcal{P}\{t,R,V,\,\chi\,\}\,\equiv \, %
\sum_{s=0}^\infty\, \frac 1{s!}\,\left\{ %
\int_1 \dots\int_s \chi(x_1) \dots \chi(x_s)\, %
\eo \prod_{j=1}^{s} %
\circ [\,\ej -1\,]\,g(x_j)\,\right\} %
\,F_0^{in}(R,V) \,\,\, \label{eve}
\end{eqnarray}
The latter equality in (\ref{gc}) directly follows from %
CF's definitions (\ref{cf}), with functional %
$\,\mathcal{P}\{t,R,V,\,\chi\,\}\,$ representing %
unified ``propagator'' for all CF's. %
In view of (\ref{fc}), %
\begin{eqnarray}
\mathcal{F}\{t,R,V,\psi\,;\,n\,\}\,=\, %
\mathcal{C}\{t,R,V,\psi\,;\,n\,\}\, %
\exp{\left\{ n\int \psi(x)\, %
g(x)\, dx\, \right\}}\,\,\, \label{fcf}
\end{eqnarray}

Since, again by the CFs definition, %
$\,C_s(t=0)=0\,$ at $\,s>0\,$ and therefore %
$\,\mathcal{C}\{t=0,R,V,\psi;n\}= %
F_0^{in}(R,V)\,$, %
one can see that the exponential in (\ref{fcf}) %
is conditional characteristic functional of %
thermodynamically equilibrium gas: %
\begin{eqnarray}
\frac {\mathcal{F}\{t=0,R,V,\psi;n\}} %
{F_0^{in}(R,V)}\,= \, %
\exp{\left\{ n\int \psi(x)\, %
g(x)\, dx\, \right\}}\,=\, %
\left\langle\, \exp{\left\{ %
\int \widetilde{n}(t,x)\, \ln\,[1+\psi(x)]\, %
dx\, \right\}}\, \right %
\rangle_{R,V,n}^{eq} \,\, \, \label{eqf}
\end{eqnarray}
Factual independence of this functional on %
$\,\{R,V\}\,$ just reflects the gas'  %
equilibrium, which is indicated by superscript ``eq''. %
The change of variable\, %
$\,\psi(x)=\exp{[\xi(x)]}-1\,$\, transforms (\ref{eqf}) into
\begin{eqnarray}
\exp{\left\{ n\int g(x)\, [\,e^{\,\xi(x)} -1\,]\, %
dx\, \right\}}\,=\, %
\left\langle\, \exp{\left\{ %
\int \widetilde{n}(t,x)\, \xi(x)\, %
dx\, \right\}}\, \right %
\rangle_{R,V,n}^{eq} \,\,, \, \label{eqf1}
\end{eqnarray}
which shows that corresponding statistics %
of random disposition of gas atoms in $\,\mu\,$-space %
is trivial Poissonian. %

However, the equilibrium of gas itself does not %
mean that of the whole system, if %
$\,F_0^{in}(R,V)\,$ differs from %
$\,\Omega^{-1}G_M(V)\,$. Then at $\,t>0\,$  %
the above mentioned excess ``historical'' correlations %
between gas and BP will appear to be accumulated by GF (\ref{gc}). %
For the first time these correlations, in their connections with %
BP's diffusion law, were considered in \cite{p0705,p0710}. %
In our present case of BP in ideal gas, %
these connections, expressed by DVRs (\ref{dvr}), together form %
generating DVR \cite{p0806,tmf,p1105} %
(quite obvious from (\ref{eve}))
\begin{eqnarray}
\mathcal{C}\{t,R,V,\,\sigma +\psi\,;\,n\,\}\,=\, %
\mathcal{C}\{t,R,V,\,\,\psi/(1+\sigma)\, %
;\,(1+\sigma)\,n\,\}\,\,, \,\, \label{gdvr} %
\end{eqnarray}
where\, $\,\sigma =\,$const\, is independent on $\,x\,$. %

\section{Diffusion law and %
accompanying gas statistics}

Combining the generating virial relation %
(\ref{gdvr}) with (\ref{gfc}) and (\ref{fcf}), %
we can write
\begin{eqnarray}
F_0(t,R,V;\,n)\, \frac {\left\langle\, \exp{\left\{ %
\int \widetilde{n}(t,x)\, \ln\,[1+\sigma +\psi(x)]\, %
dx\, \right\}}\, \right \rangle_{R,V,\,n}} %
{\exp{\left\{ n\int \sigma \, g(x)\, dx\, \right\}}} %
\,=\, \label{gcdvr}\\ \,=\, %
F_0(t,R,V;\,(1+\sigma)\,n)\, %
\left\langle\, \exp{\left\{ %
\int \widetilde{n}(t,x)\, \ln\,[1+\psi(x)/(1+\sigma)]\, %
dx\, \right\}}\, \right \rangle_{R,V,\,(1+\sigma)\,n} %
\,\, \, \nonumber
\end{eqnarray}
As the consequence, after special choice %
$\,\psi(x)=\phi(\rho)\,$ and integration %
over $\,V\,$, we have

\begin{eqnarray}
W(t,R;\,n)\, \frac {\left\langle\, \exp{\left\{ %
\int \widetilde{n}(t,\rho)\, \ln\,[1+\sigma +\phi(\rho)]\, %
d^3\rho\, \right\}}\, \right \rangle_{R,\,n}} %
{\exp{\left\{ n\int \sigma \, E(\rho)\, d^3\rho\, \right\}}} %
\,=\, \label{gwdvr} \\ \,=\, \nonumber %
W(t,R;\,(1+\sigma)\,n)\, %
\left\langle\, \exp{\left\{ %
\int \widetilde{n}(t,\rho)\, \ln\,[1+\phi(\rho)/(1+\sigma)]\, %
d^3\rho\, \right\}}\, \right %
\rangle_{R,\,(1+\sigma)\,n} \,\, \,
\end{eqnarray}
This generating DVR %
directly connects the diffusion law %
$\,W(t,R;n)=W_0(t,R;n)=\int F_0(t,R,V;n)\, d^3V\,$ with %
gas statistics in configurational space as described by one more GF %
\begin{eqnarray}
\mathcal{W}\{t,R,\phi\,;\,n\,\}\,=\, %
W_0\,+  \sum_{s=1}^\infty\, \frac {n^s}{s!} %
\int_1 \! \dots \! \int_s W_s\,\, %
\phi(\rho_1) \dots \phi(\rho_s)\, =\, \label{gw}\\ \,=\, %
W_0(t,R;n)\, \left\langle\, \exp{\left\{ %
\int \widetilde{n}(t,\rho)\, \ln\,[1+\phi(\rho)]\, %
d^3\rho\, \right\}}\, \right \rangle_{R,n} %
\,\,, \, \nonumber %
\end{eqnarray}
where now $\,\int_j \dots\, =\int \dots\, d^3\rho_j\,$.
Clearly, the integrals in both numerator and denominator %
on the left in (\ref{gcdvr}) and (\ref{gwdvr}) are %
formally diverging at $\,\sigma\neq 0\,$, %
but the divergencies definitely compensate one another, %
due to equalities (\ref{eqf})-(\ref{eqf1}), %
since at $\,\rho\rightarrow\infty\,$ random fields %
$\,\widetilde{n}(t,x)\,$ and $\,\widetilde{n}(t,\rho)\,$ %
behave like equilibrium Poissonian ones. %

Because of arbitrariness of %
$\,\sigma\,$ and $\,\psi(x)\,$ or $\,\phi(\rho)\,$, %
each of relations (\ref{gcdvr}) and (\ref{gwdvr}) produces two %
independent relations. Namely, for example, (\ref{gwdvr}) yields %
\begin{eqnarray}
\frac {W(t,R;\,(1+\sigma)\,n)} %
{W(t,R;\,n)}\,=\, %
\left\langle\, \exp{\left\{ %
\int [\,\widetilde{n}(t,\rho)\, \ln\,(1+\sigma)\,-\, %
n\,E(\rho)\,\sigma\,]\, d^3\rho\, \right\}}\, %
\right \rangle_{R,\,n} \,\,, \,\label{dls} %
\end{eqnarray}
or, writing\, $\,\sigma =\exp{(\xi)}-1\,$, %
\begin{eqnarray}
\frac {W(t,R;\,e^{\,\xi} n)} {W(t,R;\,n)}\,=\, %
\frac {\left\langle\, \exp{\left\{ %
\,\xi \int \widetilde{n}(t,\rho)\, d^3\rho\, \right\}}\, %
\right \rangle_{R,\,n}} %
{\left\langle\, \exp{\left\{ %
\,\xi \int \widetilde{n}(t,\rho)\, d^3\rho\, \right\}}\, %
\right \rangle^{eq}_{n}} \,\, \,\label{dls_} %
\end{eqnarray}
And, taking\, $\,\phi(\rho)= (1+\sigma)\, %
[\exp{(\varphi(\rho))} -1] %
= e^{\,\xi} [\exp{(\varphi(\rho))} -1]\,$, %
\begin{eqnarray}
\frac {\left\langle\, \exp{\left\{ %
\int \widetilde{n}(t,\rho)\, %
[\,\xi \,+ \, \varphi(\rho)\,]\, %
d^3\rho\, \right\}}\, \right \rangle_{R,\,n}} %
{\left\langle\, \exp{\left\{ \int %
\widetilde{n}(t,\rho)\,\xi\, %
d^3\rho\, \right\}}\, \right \rangle_{R,\,n}} %
\,=\, 
\left\langle\, \exp{\left\{ %
\int \widetilde{n}(t,\rho)\, \varphi(\rho)\, %
d^3\rho\, \right\}}\, \right %
\rangle_{R,\,e^{\,\xi}\,n} %
\,\, \, \label{sp}
\end{eqnarray}
The first of these three equalities is the same as
\begin{eqnarray}
W(t,R;\,(1+\sigma)\,n\,)\,=\, %
\mathcal{C}\{t,R,\,\sigma \,;\,n\,\} %
\,\,, \,\, \label{dls1} %
\end{eqnarray}
which follows from (\ref{gdvr}) at $\,\psi(x)=0\,$ %
after integration over BP's velocity, with GF %
\begin{eqnarray}
\mathcal{C}\{t,R,\,\phi\,;n\}\,\equiv\, %
\int \mathcal{C}\{t,R,V,\, %
\psi(x)=\phi(\rho)\,;n\}\, d^3V\,=\,
 \nonumber\\ \,=\, %
C_0(t,R;n)+\sum_{s=1}^\infty \frac {n^s}{s!} %
\int C_s(t,R,\rho^{(s)};n)\,  %
\phi(\rho_1) \dots \phi(\rho_s) \, %
d\rho^{(s)}\,=\, \frac {\mathcal{W}\{t,R,\,\phi\,;n\}} %
{\exp{\left[\,n\int E(\rho)\,\phi(\rho) %
\,d^3\rho\,\right]}}  \,\, \label{cw}
\end{eqnarray}
of CFs integrated over all velocities, %
$\,C_s(t,R,\rho^{(s)};n) \equiv %
\int\int C_s\, dv^{(s)}\, d^3V\,$.


Now, let us discuss the above approximate asymptotic expressions %
for $\,W(t,R;n)\,$ at $\,f=0\,$  %
in the light of the exact relations (\ref{dls})-(\ref{dls_}) %
or (\ref{dls1})) and (\ref{sp}). %

In the Gaussian model (\ref{g0}), %
with $\,D(n)\propto 1/n\,$, and in %
our approximation  (\ref{ng0}) left side of %
(\ref{dls}) and (\ref{dls_}) looks as %
\begin{eqnarray}
\frac {W^{(G)}_{D((1+\sigma)\,n)}(t,R)} %
{W^{(G)}_{D(n)}(t,R)}\,=\,(1+\sigma)^{3/2}\,%
\exp{(-\sigma z)}\,=\,\nonumber\\ \,=\, %
\exp{\left[\frac {3\,\xi}2 -z\, %
\left(e^{\,\xi}-1\right)\right]}\,=\, %
\frac {W^{(G)}_{D(\exp{(\xi)}\,n)}(t,R)} %
{W^{(G)}_{D(n)}(t,R)}\,\,, \,\,\label{gm}\\ %
\frac {W(t,R;\,(1+\sigma)\,n)} %
{W(t,R;\,n)}\,=\, %
\frac {(1+\sigma)^{3/2}} %
{(1+\sigma\,z/(1+z))^{\Delta N_- +3/2}} \,=\, %
\nonumber\\ \,=\, %
\exp{\left\{\frac {3\xi}2 - %
\left(\Delta N_- +\frac 32\right)\, %
\ln\,\left[\,1+ \frac z{1+z}\, %
\left(e^{\,\xi}-1\right)\right]\right\}}\,=\, %
\frac {W(t,R;\,e^{\,\xi}\,n)} %
{W(t,R;\,n)}\,\,, \,\label{ngm} %
\end{eqnarray}
\end{widetext}
where\, $\,z=R^2/4D(n)t\,\propto n\,$\, and %
$\,\sigma =\exp\,\xi\,-1\,$. %



At $\,z=0\,$  both these expressions reduce to %
$\,(1+\sigma)^{3/2}=\exp{[(3/2)\,\xi ]}\,$. %
From viewpoint of right-hand side of (\ref{dls})-(\ref{dls_})
it says that the integral %
$\, \int \widetilde{n}(t,\rho)\, d^3\rho\,\equiv \widetilde{N}(t)\,$, - %
representing number of atoms in some relevant volume %
around BP, - consists of non-random constant $\,3/2\,$  %
and random component which behaves exactly like %
$\, \widetilde{N}(t)\,$ in equilibrium, with %
the same Poissonian statistics. Symbolically, %
$\, \widetilde{N}(t)= 3/2 + %
\widetilde{N}_{eq}(t)\,$\,.

Literal interpretation of this statement would mean  %
that, strangely, $\, \widetilde{N}(t)\,$ takes non-integer values. %
However, such interpretation would be ill-advised, %
since the proportionality %
$\,D(n)\propto 1/n\,$, - which just causes factor %
$\,(1+\sigma)^{3/2}=\exp{[(3/2)\,\xi ]}\,$ (through %
$\,W(t,0;n)\propto (4\pi D(n)t)^{-3/2}\,$), - %
is approximate and applicable only when %
\begin{eqnarray}
(\pi a^2 V_0t)^{-1}\,\ll\, n\,\ll\, %
(4\pi a^3/3)^{-1} \,\, \, \nonumber
\end{eqnarray}
and, besides,\, $\,(1+\sigma)\,n =\exp{(\xi)}\,n\,$ also %
lies within these bounds. %
What we can state, is that contribution $\,3/2\,$ in %
$\, \widetilde{N}(t)= 3/2 + \widetilde{N}_{eq}(t)\,$ %
indicates BP-atoms statistical correlations %
produced, of course, by (actual or virtual) %
collisions of atoms with BP. %
At  $\,z=0\,$, in that way, that are %
naturally positive correlations: if in spite of $\,t\gg \tau\,$ %
BP has not gone away far from its start %
position, then  number of atoms in its surroundings %
on average is greater than under initial equilibrium %
(for fixed BP), so that
\begin{eqnarray}
\overline{\Delta N}(t|0,n)\,=\, %
\langle\widetilde{N}(t)\rangle_{R=0,n}\,-\, %
\langle\widetilde{N}_{eq}\rangle_{n}\,=\, %
3/2 \,\, \label{r0} %
\end{eqnarray}
with $\,\langle\widetilde{N}_{eq}\rangle_{n}= %
\langle\widetilde{N}(t)\rangle_{n}^{eq}\,$.

\,\,\,

At finite $\,z>0\,$, obviously, in both the cases (\ref{gm}) and (\ref{ngm}) %
$\,\overline{\Delta N}(t|R,n)\,$ decreases and becomes %
negative at $\,z>3/2\,$ and $\,z>3/2\Delta N_-\,$, respectively, %
again in agreement with intuition: %
the greater is current BP's distance from its start point, %
the smaller is number of atoms around it, i.e. %
``number of obstacles'' to its flight,  in comparison with %
equilibrium. But numerical and statistical characteristics %
of this atoms' ``shortage'' in cases (\ref{gm}) and (\ref{ngm}) %
are qualitatively different. %

Namely, the Gaussian model, according to (\ref{gm} or (\ref{gdvr1}), %
must be associated with infinitely growing shortage, %
$\,\overline{\Delta N}(t|R,n)\,=\,3/2 -z\,$. %
Thus, if $\,|R|\sim V_0t\,$  then the shortage is of order %
of $\,z\sim t/\tau\,$, i.e. as large as a number of %
missed BP-atom collisions,  as if long ago missed atoms %
somehow determined current shortage. %
Indeed, this is physically rather absurd picture! %

Another picture, which arises under our approximation, looks %
much more likely. Here, according to (\ref{neq}) and (\ref{ngm}), %
mean shortage is bounded,
\begin{eqnarray}
\overline{\Delta N}= %
\frac 32 - \left[\Delta N_- + \frac 32\right]\,\frac z{1+z}\, %
\rightarrow\, -\Delta N_- \,\, \, \label{rinf} %
\end{eqnarray}
Besides, it obeys essentially non-Poissonian statistics, %
which means presence of statistical correlations between composing %
atoms. %
One may describe this statistics by treating %
the ``microscopic gas density'' %
$\,\widetilde{n}(t,\rho)\,$ as ``twice stochastic'' %
point field (``spatial random point process''), i.e. %
Poissonian field with fluctuating intensity. %
Then, applying in (\ref{ngm}) integral expansion
\begin{eqnarray}
\exp{\left\{ - \left(\Delta N_- +\frac 32\right)\, %
\ln\,\left[\,1+ \frac z{1+z}\, %
\left(e^{\,\xi}-1\right)\right]\right\}}\,=\, %
\nonumber\\ \,=\, %
\int_0^{\infty} \exp{\left[ - u\, %
\frac z{1+z}\, \left(e^{\,\xi}-1\right)\right]}\, %
\times \nonumber\\ \times\,\, %
\frac {u^{\,\Delta N_- + 1/2}\,\exp{(-u)}} %
{\Gamma(\Delta N_- +3/2)}\,\, du\,\,, \,\,\,\,\,\,\,\, \label{nge} %
\end{eqnarray}
we can treat factor  %
\begin{eqnarray}
\Delta N\,\equiv\, 3/2\,-\, %
u\,z/(1+z)\,\, \, \label{rdn}
\end{eqnarray}
(with $\,u\,$ obeying the gamma distribution) %
as fluctuating excess or shortage (dependently on %
$\,\Delta N\,$'s sign) of number of atoms  %
in correlation volume near BP. At that, its average value %
equals to (\ref{rinf}) %
while most probable value, %
$\,\Delta N_{m.p.}(z)\,$, is smaller, %
\begin{eqnarray}
\Delta N_{m.p.}= \frac 32-\left[\Delta N_- + %
\frac 12\right] \frac z{1+z}\, %
\rightarrow\,1-\Delta N_-\, \,\,\, \label{mpd}
\end{eqnarray}
(since most probable value of $\,u\,$ is %
$\,\Delta N_- +1/2\,$).


Let us use it for estimation %
of limits in (\ref{rinf}) and (\ref{mpd}), that is %
characteristic correlation %
number $\,\Delta N_-\,$. Introduce quantity %
\begin{eqnarray}
\Delta M\,\equiv\, M\,+\,m\,\Delta N\,\, \label{dm}
\end{eqnarray}
It represents excess (or shortage) of total mass %
(of BP and atoms) located inside the correlation volume %
$\,\Omega_c\,$, in comparison with mean mass %
$\,mn\Omega_c\,$ of other regions having %
such volume but not containing BP. %
Notice, first, that in perfectly equilibrium statistical ensemble, -  %
where BP's position is fully uncertain,   %
$\,W(R)=1/\Omega\,$, - the same quantity $\,mn\Omega_c\,$
gives mean mass of arbitrary region with volume $\,\Omega\,$, %
even though it may contain BP. %
Second, since in our  ensemble the limit value of %
$\,\Delta N\,$ at $\,z\rightarrow\infty\,$ appears insensible %
to $\,z\,$,\, it in fact also corresponds to  fully uncertain %
BP's position (``at infinity''). %
Therefore we can expect that the excess-shortage %
mass $\,\Delta M\,$ in our ensemble %
at $\,z\rightarrow\infty\,$ becomes the same as in perfectly equilibrium %
ensemble, i.e. turns to zero. %
Then, if addressing this requirement to most probable value %
of $\,\Delta M\,$, i.e. identifying $\,\Delta N\,$ in (\ref{dm}) with %
$\,\Delta N_{m.p.}(\infty)\,$ from (\ref{mpd}), we come to equality %
\begin{eqnarray}
M +(1-\Delta N_-)\,m\,=\,0\,\,, \, \label{dm0}
\end{eqnarray}
which suggests for $\,\Delta N_-\,$ value\, %
$\,\Delta N_- =1+M/m\,$. %

\,\,\,

This value coincides with what was found formally, %
under the collisional approximation, in \cite{p1,p2} %
for BP in non-ideal gas. Of course, our derivation of (\ref{dm0}) %
is less formally grounded, and factual value %
of $\,\Delta N_-\,$ may differ from $\,1+M/m\,$. %
Nevertheless, undoubtedly, this result %
correctly reflects role of mass ratio %
$\,M/m\,$ in construction of the law of diffusion, and %
confirms $\,\Delta N_-\,$'s indifference to gas density.

\,\,\,

As the consequence, we can %
improve estimate of the correlation volume $\,\Omega_c\,$ %
and make estimate of its analogues for higher-order CFs. %
With this purpose let us return to formula (\ref{ir}).  %
Due to relations (\ref{fc}) we can write %
\begin{eqnarray}
\overline{W}_s = %
\sum_{k=0}^s \left( \begin{array}{c} %
s \\ k \end{array} \right)\, %
\overline{C}_k \,\,, \,  \nonumber 
\end{eqnarray}
where $\,\overline{C}_s =\overline{C}_s(t,R;n)\,$ %
are CFs smoothed over small enough %
$\,\rho^{(s)}\,$. Combining this with (\ref{ir}) and %
DVR (\ref{dvr}), we obtain %
\begin{eqnarray}
\overline{C}_s \,=\, %
\frac{\Gamma(\Delta N_-)} %
{\Gamma(\Delta N_- +s)}\, %
\,n^s \int_1 \dots \int_s C_s\,\, \label{mcf}
\end{eqnarray}
The factor before integral here just has meaning of inverse %
$\,3s\,$-dimensional characteristic volume occupied by %
$\,(s+1)$-order irreducible correlation between BP and %
$\,s\,$ atoms. Thus the $\,3s\,$-dimensional volume itself is %
\begin{eqnarray}
\Omega_{c\,s}\,=\, \frac %
{\Delta N_-\,\dots\, (\Delta N_- +s-1)} %
{n^s}\,\,, \,   \label{cv}
\end{eqnarray}
in particular, $\,\Omega_{c}\equiv \Omega_{c\,1}=\Delta N_-/n\,$. %

This expression, being supplemented with above estimate %
$\,\Delta N_- =1+M/m\,$, acquires visual explanation. %
Namely, at $\,M/m \lesssim 1\,$ we see from (\ref{cv}) that\, %
$\,\Omega_{c\,s}^{1/s} \sim ((s+1)!)^{1/s}/n %
\sim (s+1)/en\,$\,. Hence, 3D correlation volume %
of a particular BP-atom link (in $\,\rho_j\,$-space) %
grows proportionally to $\,s\,$, that is %
correlation between BP and one of atoms is mediated %
and lengthened by all $\,s-1\,$ others. Physically, this means that %
all they are participants of a same connected cluster %
of $\,s\,$ (actual or virtual) collisions %
(or ``encounters'' \cite{i1,i2}), and most  %
long-range contributions to $\,\Omega_{c\,s}^{1/s}\,$ %
come from cylinder-like regions (``collision cylinders'') with %
cross-section $\,\sim \pi a^2\,$ and length\, $\,\sim s\,\lambda\,$  %
(thus, with volume $\,\sim s\pi a^2\lambda = s/n\,$). %

\,\,\,

If $\,M/m \gg 1\,$, then formula (\ref{cv}) implies, %
at not too large $\,s\,$, value\, $\,\Omega_{c\,s}^{1/s}\sim %
\Omega_c \sim M/mn\,$\,. Its approximate independence on $\,s\,$ %
says about relative weakness of (BP-mediated) inter-atom %
correlations inside the clusters (if BP was immovable, %
they would vanish at all). %

Physical meaning of the proportionality\, %
$\,\Omega_c \propto M/m\,$\, at $\,M/m \gg 1\,$ %
also is quite transparent.  Indeed, $\,\Omega_c\,$ is %
formed by ``collision cylinders'' with length $\,\Lambda \,$ of order %
of atom's flight path during BP's velocity relaxation %
time $\,\tau\,$, i.e. $\,\Lambda \sim \tau \sqrt{T/m}\,$. %
At that,\, $\,\tau\sim $ $M\lambda/\sqrt{Tm}\,\,$\, %
\cite{pk}. %
This value follows e.g. from phenomenological equation %
for BP's momentum relaxation in light dilute gas:\, %

\,\,\,\,\,\, $\,M\dot{V}\approx $ $- (V/\lambda)\sqrt{Tm}\,$\,. %
\, Hence, $\,\Lambda \sim (M/m)\,\lambda\,$, %
thus explaining why\, %
$\,\Omega_c =\Delta N_-/n \approx \pi a^2 \Lambda  \approx M/mn\,$.

\,\,\,

Above reasonings demonstrate that characteristic ``correlation %
volume'' can be thought as a weighted sum (union), or statistical sum, %
of variously oriented (and may be coupled) ``collision cylinders''. %
That is why volume $\,\Omega_c\,$ is determined by $\, 1/n\,$, %
but not $\, \lambda^3\,\gg 1/n\,$, although its %
greatest liner size is determined by $\, \lambda\,$. %
Such construction of the correlations implies strongly non-uniform %
distribution of ``density of correlations'' inside  $\,\Omega_c\,$. %
For instance, - as was suggested already in \cite{p0710} %
(and discussed also in \cite{p1203}), - schematically,\, %
$\,C_1(t,R,\rho;n)/\overline{C}_1(t,R;n)\sim $ %
$(a^2/4|\rho|^2)\,\exp{(-|\rho|/\Lambda)}\,$\, %
at $\,|\rho|\gtrsim a\,$,\, which reveals hidden motive of %
relations (\ref{mcf})-(\ref{cv}). At this point, however, we enter %
to ``terra incognita''.

\section{Conclusion}

We have tested one more novel and unprejudiced approach %
to the problem about molecular Brownian particle (BP) in ideal gas, %
basing on explicit density expansions of %
(time-depending non-equilibrium) probability %
distribution functions and on exact dynamical virial relations (DVR) %
obtained directly from these expansions. %
Previous approaches were based on the ``generalized %
fluctuation-dissipation relations'' (FDR) \cite{p0802,p0803,p0710} %
(about FDR themselves and their other applications see %
e.g. \cite{bkn} and references therein),\, on the BBGKY equations %
and Bogolyubov's generating functional equation %
\cite{p0806,tmf,ig,p1105,p0804,p0710},\, on the ``stochastic representation %
of deterministic interactions'' \cite{sr} and path %
integrals \cite{p0806,sr}.\, %
So much attention to seemingly too particular problem was caused %
by understanding that in fact it is very non-trivial for real ``honest'' %
statistical mechanics (SM). Though it is quite trivial for  %
``Boltzmannian version'' of SM %
exploiting ``Bernoullian way of thinking'', i.e. belief that %
many-particle chaos can be divided into ``independent'' elementary %
events with strictly certain probabilities.  %
In the framework of this fantasy, long-range statistics of random walk %
of molecular-size BP does not differ from %
standard mathematical Brownian motion (diffusive process, %
Wiener process, etc.), while in real SM one reveals %
random walk without certain diffusivity and mobility or, in other %
words, with 1/f-type fluctuations of diffusivity and mobility. %
This difference, firstly, %
illustrates Krylov's statement \cite{kr} %
that the only ``elementary events'' what generally %
can be independent and have %
certain probabilities  are whole phase trajectories %
of one or another many-particle system. %
Secondly, highlights where one should search  for %
origin of various 1/f noises. %
They, along with accompanying historical correlations \cite{tmf}, %
manifest uniqueness of any factual phase %
trajectory (experiment). %
Thirdly, the mentioned difference %
once again shows that %
``surprises in theoretical physics'' %
do continue, and failure of the ``law of conservation of %
slopiness'' \cite{rp} (in molecular chaos %
considerations) supplements their collection.
Hence, the problem under our attention is of principal importance. %

\,\,\,

We developed analysis of %
our present system in terms of %
conditional statistical characteristics of  %
microscopic gas density under given BP's path passed during total %
observation time. We transformed the DVR into %
simple and  intelligible relations between these %
gas characteristics and BP's diffusion law. First of all, %
principal relation between its scaling exponents in respect to mean gas density %
and BP's path value, on one hand, and %
conditional mean number of gas atoms %
covered by instant statistical correlations with BP, %
on the other hand. We demonstrated that such relations   %
predict essentially non-Gaussian BP's %
law of diffusion possessing nearly power-law long tail. %
Moreover, they, - in combination with natural heuristic %
reasonings, - allow quantitative estimate of %
the tail's exponent even without formal summation %
of the explicit expansions.

The ancient classical ``kinetic theory of gases'' is unfamiliar %
with so remarkable relations, but now gets %
chance to enhance its ``mental outfit'' and language  %
and include notion of molecular chaos %
which produces 1/f noise. %
Simultaneously, of course, we need in %
more complete mathematical investigation (if not exact solution) %
of the problem about BP in ideal gas, along with %
other problems about spatial-temporal %
statistics of relaxation and transport processes %
in many-particle systems.


%



\begin{thebibliography}{24}
\bibitem{fn}
It is necessary to notice that formula before formula %
[17] in \cite{p2}  contains two misprints: %
the letter $\,n\,$ under sums over $\,s\,$ in %
factors $\,B(c,n)\,$ and $\,t^{n-1}\,$ should be replaced %
by letter $\,s\,$.
\end{thebibliography}
\end{document}